\documentclass{pnastwo}

\usepackage{amssymb,amsfonts,amsmath}
\usepackage{graphicx}
\usepackage{lscape}
\usepackage{booktabs}
\usepackage{multirow}
\usepackage{color}

\contributor{Submitted to Proceedings
of the National Academy of Sciences of the United States of America}
\url{www.pnas.org/cgi/doi/10.1073/pnas.0709640104}
\copyrightyear{2008}
\issuedate{Issue Date}
\volume{Volume}
\issuenumber{Issue Number}

\begin{document}



\title{Collective credit allocation in science}





\author{Hua-Wei Shen\affil{1}{Institute of Computing Technology, Chinese Academy of Sciences, Beijing 100190, China}\affil{2}{Center for Complex Network Research, Department of Physics, Department of Biology, and Department of Computer Science, Northeastern University, Boston, MA 02115, USA},
Albert-L\'aszl\'o Barab\'asi\affil{2}{}\affil{3}{Center for Cancer Systems Biology, Dana-Farber Cancer Institute, Boston, MA 02115, USA}\affil{4}{Department of Medicine, Brigham and Women's Hospital, Harvard Medical School, Boston, MA 02115, USA}\affil{5}{Center for
Network Science, Central European University, Budapest 1051, Hungary}}

\contributor{Submitted to Proceedings of the National Academy of Sciences
of the United States of America}

\maketitle

\begin{article}

\begin{abstract}
Collaboration among researchers is an essential component of the modern scientific enterprise, playing a particularly important role in multidisciplinary research. However, we continue to wrestle with allocating credit to the coauthors of publications with multiple authors, since the relative contribution of each author is difficult to determine. At the same time, the scientific community runs an informal field-dependent credit allocation process that assigns credit in a collective fashion to each work. Here we develop a credit allocation algorithm that captures the coauthors' contribution to a publication as perceived by the scientific community, reproducing the informal collective credit allocation of science. We validate the method by identifying the authors of Nobel-winning papers that are credited for the discovery, independent of their positions in the author list. The method can also compare the relative impact of researchers working in the same field, even if they did not publish together. The ability to accurately measure the relative credit of researchers could affect many aspects of credit allocation in science, potentially impacting hiring, funding, and promotion decisions.
\end{abstract}






\section{Significance Statement}
The increasing dominance of multi-author papers is straining the credit system of science: while for single-author papers the credit is obvious and undivided, for multi-author papers credit assignment varies from discipline to discipline. Consequently each research field runs its own informal credit allocation system, which is hard to decode for outsiders. Here we develop a discipline-independent algorithm to decipher the collective credit allocation process within science, capturing each coauthor's perceived contribution to a publication. The proposed method provides scientists and policy-makers an effective tool to quantify and compare the scientific contribution of each researcher without requiring familiarity with the credit allocation system of the specific discipline.

\section{Introduction}

Reflecting the increasing complexity of modern research, in the past decades collaboration among researchers became a standard path to discovery~\cite{Wuchty2007}. Collaboration plays a particularly important role in multidisciplinary research that requires expertise from different scientific fields~\cite{Lawrence2007}. As the number of coauthors of each publication increases, science's credit system is under pressure to evolve~\cite{Hodge1981,Kennedy2003,Allen2014}. For single-author papers, which were the norm decades ago, credit allocation is simple: the sole author gets all the credit. This rule, accepted since the birth of science, fails for multi-author papers~\cite{Sekercioglu2008}. The lack of a robust credit allocation system that can account for the discrepancy between researchers' contribution to a particular body of work and the credit they obtain, has prompted some to state that ``multiple authorship endangers the author credit system''~\cite{Greene2007}. This situation is particularly acute in multidisciplinary research~\cite{Biggs2008,Kaur2013}, when communities with different credit allocation traditions collaborate~\cite{Lehmann2006}. Furthermore, a detailed understanding of the rules underlying credit allocation is crucial for an accurate assessment of each researcher's scientific impact, affecting hiring, funding, and promotion decisions.

Current approaches to allocating scientific credit fall in three main categories. The first views each author of a multi-author publication as the sole author~\cite{Garfield1972,Hirsch2005}, resulting in inflated scientific impact for publications with multiple authors. This system is biased towards researchers with multiple collaborations or large teams, customary in experimental particle physics or genomics. The second assumes that all coauthors contribute equally to a publication, allocating fractional credit evenly among them~\cite{Egghe2006,Hirsch2007}. This approach ignores the fact that authors' contributions are never equal, hence dilutes the credit of the intellectual leader. The third allocates scientific credit according to the order or the role of coauthors, interpreting a message agreed upon within the respective discipline~\cite{Hagen2008,Zhang2009,Stallings2013}. For example, in biology typically the first and the last author(s) get the lion's share of the credit and in some areas of physical sciences the author list reflects a decreasing degree of contribution. An extreme case is offered by experimental particle physics where the author list is alphabetic, making it impossible to interpret the author contributions without exogenous information. Finally, there is an increasing trend to allocate credit based on the specific contribution of each author~\cite{Tscharntke2007,Clippel2008} specified in the contribution declaration required by some journals~\cite{Foulkes1996,Campbell1999}. Yet, each of these approaches ignores the most important aspect of credit allocation: notwithstanding the agreed upon order, credit allocation is a collective process~\cite{Radicchi2009,Rybski2009,Mones2014}, which is determined by the scientific community rather than the coauthors or the order of the authors in a paper. This phenomena is clearly illustrated by the 2012 Nobel prize in physics that was awarded based on discoveries reported in publications whose last authors were the laureates~\cite{Brune1996,Meekhof1996}, while the 2007 Nobel prize in physics was awarded to the third author of a nine-author paper~\cite{Baibich1988} and the first author of a five-author publication~\cite{Grunberg1986}. Clearly the scientific community operates an informal credit-allocation system that may not be obvious to those outside of the particular discipline.

The leading hypothesis of this work is that the information about the informal credit allocation within science is encoded in the detailed citation pattern of the respective paper and other papers published by the same authors on the same subject. Indeed, each citing paper expresses its perception of the scientific impact of a paper's coauthors by citing other contributions by them, conveying implicit information about the perceived contribution of each author. Our goal is to design an algorithm that can capture in a discipline-independent fashion the way this informal collective credit allocation mechanism develops.

\section{Results}

We start by examining the simplest situation: given a paper $p_0$ with two authors, $a_1$ and $a_2$, who gets the credit? Consider the extreme case when author $a_1$ has published several other papers on the topic of paper $p_0$ that are often cited together with $p_0$; for author $a_2$, the target paper is his only publication. Given that $a_1$ has a track record in the particular discipline and $a_2$ is unknown to the community, the community views $p_0$ as a part of $a_1$'s body of work (Fig.~\ref{fig:cases}a). The credit allocation system should recognize this and assign most or all credit to $a_1$. The other extreme case is when {\it all} papers pertaining to the topic of $p_0$ are joint publications between $a_1$ and $a_2$. Lacking any exogenous information, the two authors share equal credit for the target paper (Fig.~\ref{fig:cases}b), a symmetry that should be captured by a credit allocation method. In practice the situation is more complicated: authors $a_1$ and $a_2$ may publish some papers together and several with other coauthors on the topic of $p_0$. Hence their credit share of the particular work diverges with time, based on the impact of the body of work they publish separately. Next we describe a method that can account for this collective credit allocation process.

\subsection{Credit allocation algorithm}
\begin{figure}[t]
\begin{center}
    \includegraphics[width = 0.49 \textwidth]{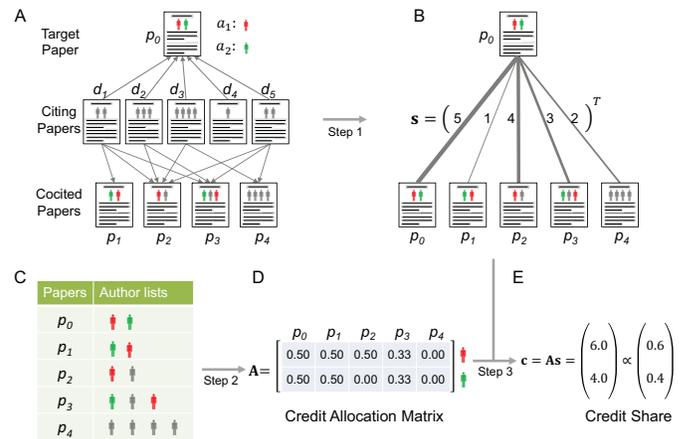}
\caption{\label{fig:method}\textbf{Illustrating the credit allocation process.} \textbf{a}, The target paper $p_0$ has two authors, $a_1$ and $a_2$, colored in red and green respectively. We also show the citing papers $d_k\, (1\leq k \leq 5)$ and the co-cited papers $p_j\, (0 \leq j \leq 4)$ that were cited by these citing papers together with $p_0$. \textbf{b},
The $p_0$-centric co-citation network constructed from \textbf{a}, where the weights of links denote the co-citation strength $s$ between the  co-cited papers and the target paper $p_0$. \textbf{c}, The author lists of the target paper $p_0$ and its co-cited papers. \textbf{d}, The credit allocation matrix $A$ obtained from the author lists of the co-cited papers in \textbf{c}. The matrix $A$ provides for each co-cited paper the authors' share. For example, since $p_2$ has $a_1$ as one of its two authors but it lacks the author $a_2$, it votes $0.5$ for author $a_1$ and $0.0$ for author $a_2$. \textbf{e}, With the matrix $A$ and co-citation strength $s$, the credit share of the two authors of $p_0$ is computed according to Eq.~[\ref{eq:credit_share}] or Eq.~[\ref{eq:matrix_form}] with a normalization.}
\end{center}
\vskip 0.15in
\end{figure}

Consider a paper $p_0$ with $m$ coauthors $\{a_i\}\,(1\leq i\leq m)$. To determine the credit share of each author, we first identify all papers that cite $p_0$, forming a set $\mathcal{D}\equiv \{d_1,d_2,\cdots,d_l\}$. Next we identify all co-cited papers $\mathcal{P}\equiv\{p_0,p_1,\cdots,p_n\}$, representing the complete set of papers cited by papers in the set $\mathcal{D}$. The relevance of each co-cited paper $p_{j}$ $(0\leq j\leq n)$ to the target paper $p_0$ is characterized by its co-citation strength $s_{j}$ between $p_0$ and $p_j$, defined as the number of times $p_0$ and $p_j$ are cited together by the papers in $\mathcal{D}$~\cite{Small1973}. For example, for $p_1$ in Fig.~\ref{fig:method}a we have $s_1=1$ since only one paper ($d_1$) cites $p_0$ and $p_1$ together, while $s_2=4$ as four papers ($d_1, d_2, d_3, d_5$) cite $p_0$ and $p_2$ together. Co-citation strength captures the intuition that papers by an author that are perceived to be very relevant to paper $p_0$ should increase the author's perceived contribution to $p_0$. Note that the target paper $p_0$ is also viewed as a co-cited paper of itself with co-citation strength equal to the citation count of $p_0$. Consequently for papers with high citation count the credit share of coauthors is less likely to be affected by other co-cited papers.

\begin{figure}[t]
\begin{center}
    \includegraphics[width = 0.42 \textwidth]{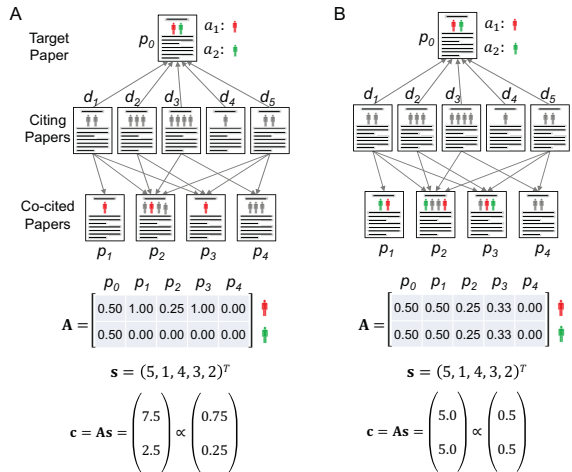}
\caption{\label{fig:cases}\textbf{Extreme cases of credit allocation.} \textbf{a}, {\it Asymmetric credit}: When author $a_2$ contributes to only one paper in a body of work, the community assigns credit to $a_1$, who publishes multiple papers on the topic. \textbf{b}, {\it Symmetric credit}: When authors $a_1$ and $a_2$ publish all their papers on the topic of paper $p_0$ jointly, they equally share the credit. In both cases, $p_0$ is the target paper with two authors $a_1$ and $a_2$ colored in red and green respectively; $d_k\, (1\leq k \leq 5)$ are citing papers of $p_0$; $p_j\, (0 \leq j \leq 4)$ are papers that were co-cited by the papers that cite $p_0$; $\bf{A}$ is the credit allocation matrix; $\bf{s}$ depicts the co-citation strength between co-cited papers and target paper, and $\bf{c}$ is the final credit share for the authors of the target paper $p_0$.}
\end{center}
\end{figure}

Using the author list of the co-cited papers, we next calculate a credit allocation matrix ${\bf A}$, whose element $A_{ij}$ denotes the amount of credit that author $a_i$ gets from co-cited paper $p_j$ (see SI Appendix: Section S2.2). To develop a discipline-independent method for credit allocation, we use a fractional credit allocation matrix that does not depend on the order of authors in the author list. For example, paper $p_1$ assigns all credit to author $a_1$ who is the sole author of $p_1$, while $p_0$ assigns equal (half) credit to authors $a_1$ and $a_2$ (Fig.~\ref{fig:cases}a). The total credit $c_i$ of author $a_i$ is the weighted sum of its local credit obtained from all co-cited papers
\begin{equation}
c_i = \sum_{j}{A_{ij}s_j},\label{eq:credit_share}
\end{equation}
or in the matrix form
\begin{equation}
\bf{c}=\bf{A}\bf{s}.\label{eq:matrix_form}
\end{equation}
The vector ${\bf c}$ provides the credit of all authors of target paper $p_0$. By normalizing ${\bf c}$ we obtain the fractional credit share among coauthors (Fig.~\ref{fig:method}e).

We apply the proposed procedure to the two extreme cases of Fig.~\ref{fig:cases}. When author $a_2$ has only one paper on the topic of $p_0$, the fact that the community cites $p_0$ together with other papers of author $a_1$ indicates that they perceive $p_0$ a part of a larger body of work by author $a_1$ (Fig.~\ref{fig:cases}a). Our method in this case obtains ${\bf c}=(0.75,\,0.25)^T$, hence allocating most credit to author $a_1$. When all subsequent work are joint, it gives ${\bf c}=(0.5,\,0.5)^T$, i.e., credit is equally shared between $a_1$ and $a_2$ (Fig.~\ref{fig:cases}b).

\begin{figure*}[tb]
\vskip 0.1in
\begin{center}
\label{fig:prediction}
\includegraphics[width = 0.98 \textwidth]{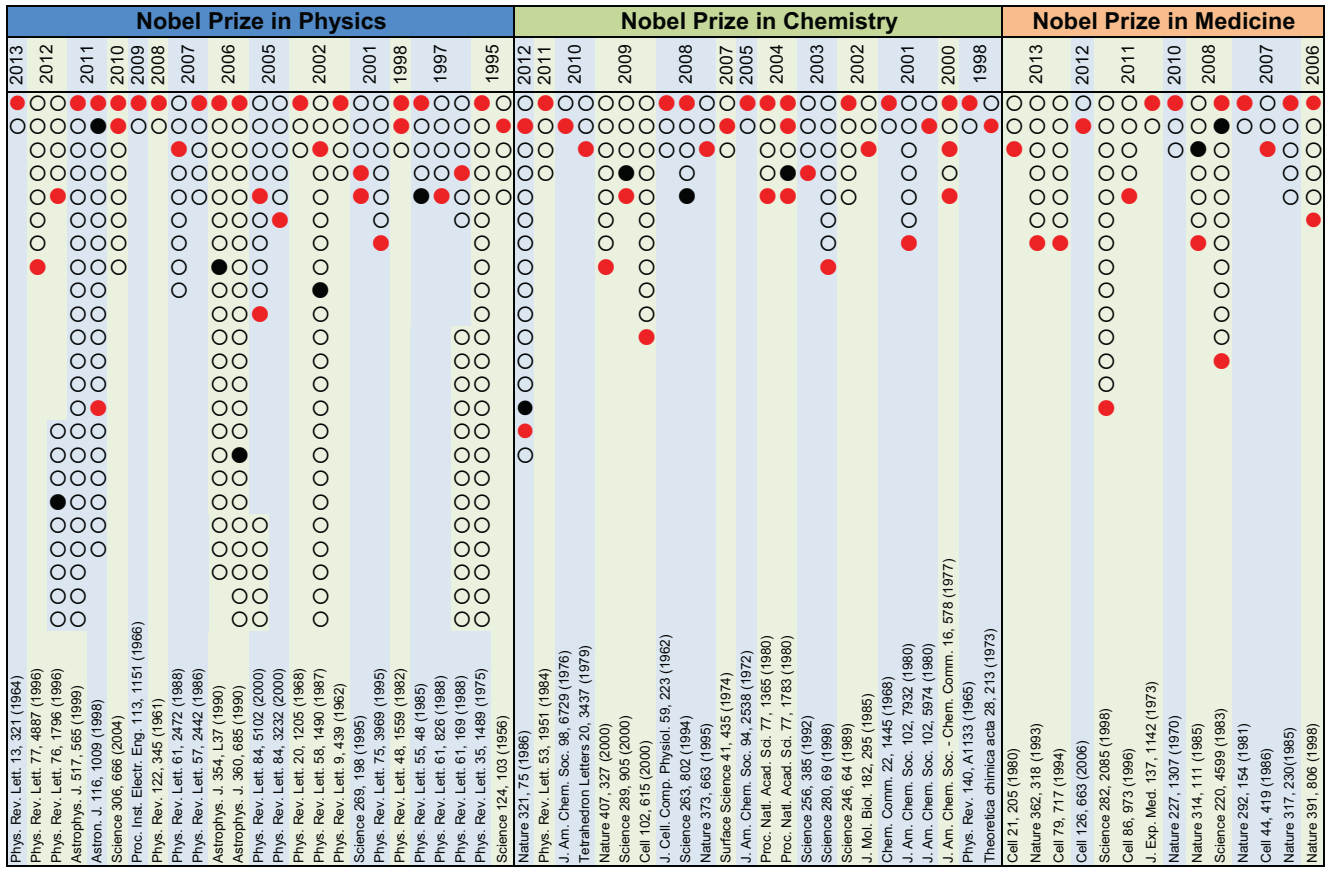}
\end{center}
\caption{\textbf{Identifying Nobel laureates from the prize-winning papers in Physics, Chemistry, and Medicine.} We apply our method to all multi-author Nobel prize-winning papers in Physics (1995--2013), Chemistry (1998--2013), Medicine (2006--2013), covering the periods since the Nobel committee started offering a detailed explanation with references for the prize. For each Nobel prize-winning paper, the laureates are shown in red-filled circles. The author with top credit share (for a paper with $k$ laureates we consider all the top-$k$ credit share) is shown as a black-filled circle when he/she is not a laureate. Other coauthors are shown as empty circles. Hence the presence of black-filled circles indicates that the credit allocation offered by our algorithm is inconsistent with the decision made by the Nobel committee. The individuals to whom we assign the top credit correspond to laureates in $51$ of the $63$ prize-winning papers. To accommodate papers with more than $23$ authors, we put in the adjacent column the circles corresponding to the authors after the $23$rd one, forming irregular blocks. Results are based on the Web of Science dataset. Papers on Economics are not shown since they are either single-author papers or are not contained in our dataset.}
\end{figure*}

\subsection{Validation}
To validate our method we apply it to Nobel prize-winning publications, representing a case where the community (and the Nobel committee) has decided where the main credit goes. We therefore collected all Nobel prize-winning papers in Physics (1995--2013), Chemistry (1998--2013), Medicine (2006--2013), and Economics (1995--2013), since the Nobel committee started offering a detailed explanation with references for the prize. Table~\ref{tab:example} shows the obtained credit share for five Nobel prize-winning physics papers in the year before the Nobel prizes were awarded, hence discounting the influence of the prize (see SI Appendix: Table S5--S8 for the complete set of results). We find that in four of the five cases the laureates have the largest credit share, no matter whether they are the first (2010) or the last authors (2012) or occupy some intermediate position in the author list (2007). For example, as the third author of the prize-winning paper with nine coauthors~\cite{Baibich1988}, the 2007 Nobel laureate A. Fert gets nearly one fourth of all credit and the remaining credit is almost evenly distributed among the other coauthors. A particularly interesting case is the 2010 prize-winning paper~\cite{Novoselov2004}, where two of eight coauthors were awarded the Nobel prize, consistent with the predicted credit share. Indeed, the credit share of the laureates is almost equal and it is 2.5 times higher than the credit share of the third-ranked coauthor. Another interesting example, out of the validation sample, is offered by the 1974 Nobel prize in Physics awarded to A. Hewish for the discovery of pulsars~\cite{Hewish1968}, with S. J. Bell as the second of five authors. Researchers in the community occasionally refer to the 1974 Nobel prize as the ``No-Bell'' prize because many feel that Bell should have shared it (http://en.wikipedia.org/wiki/Antony\_Hewish). Applying our method to the prize-winning paper, we obtain ${\bf c}=[0.250, 0.189, 0.196, 0.185, 0.180]^T$, assigning the largest credit to the laureate, indicating that the committee's choice was consistent with the perceived credit within the scientific community. Figure~\ref{fig:prediction} shows the accuracy of our method at identifying the laureates from the author list of all the $63$ multi-author prize-winning papers across three disciplines. We find that the authors with top credit share correspond to laureates in $51$ papers (81\%), despite the diversity of positions the laureates had in the author list. Note that we did not count single-author papers, for which credit is obvious. Counting those as well, accuracy increases to 86\%.

Finally, it is useful to understand potential reasons for the method's occasional failure. For example, for the two prize-winning papers of the 2011 Physics Nobel our method fails to correctly identify the laureates, caused by the fact that one researcher (Filippenko) coauthors both prize-winning papers but is not the intellectual leader for either of them. Consequently he gets the top credit on both papers. The laureates get the highest credit among the remaining coauthors, hence if the anomaly is removed, our method correctly identifies them. This case could be corrected by incorporating contextual or exogenous information into the credit allocation matrix, like the order of the authors, as we discuss below. Another fascinating anomaly is the 1997 Nobel prize in Physics~\cite{Chu1985}: S. Chu was awarded the prize although A. Ashkin has the highest credit share according to our method. Considered by many scientists the father of the field of optical tweezers~\cite{McGloin2010}, Ashkin published several high-impact papers~\cite{Ashkin1970,Gordon1980} preceding the collaboration with Chu, developing the technology that made the Nobel prize-winning discovery possible. The prize-winning paper is repeatedly co-cited with the preceding papers, explaining Ashkin's higher score. As we show below, credit to Chu is restored if we restrict the co-cited pool to papers published after the joint 1985 (Nobel-winning) paper, removing the influence of the preceding work.

\subsection{Credit share evolution}
\begin{figure}[t!]
\begin{center}
\includegraphics[width = 0.48 \textwidth]{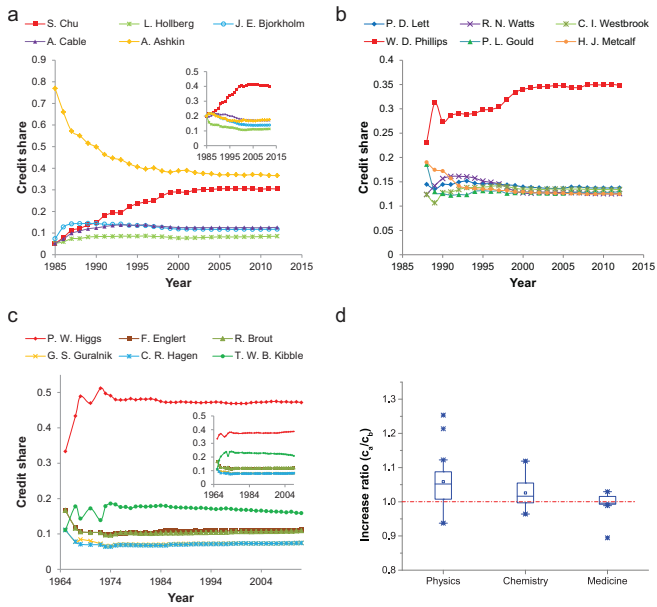}
\caption{\label{fig:evolution} \textbf{Credit share evolution.} \textbf{a}, The credit share of the authors of the 1997 Nobel prize-winning paper~\cite{Chu1985}. (Inset) Credit share obtained when we only consider the co-cited papers published after the publication of the prize-winning paper. \textbf{b}, The credit share of the authors of the other 1997 Nobel prize-winning paper~\cite{Lett1988}. \textbf{c}, Credit share of six physicists who contributed to the theory of Higgs boson in 1964~\cite{Englert1964,Higgs1964,Guralnik1964}, obtained by our method using the Web of Science dataset as input. (Inset) The same as in \textbf{c}, but based on the APS dataset. \textbf{d} The influence of Nobel prize on laureates' credit share. For each laureate we use the increase ratio ($c_a/c_b$) to quantify the change of her credit share after she was awarded the Nobel prize. In the box-plot, whiskers are higher than 90th percentile or lower than 10th percentile. Results are based on the Web of Science dataset.}
\end{center}
\end{figure}

The proposed methodology also allows us to determine the temporal evolution of credit share between coauthors. To illustrate this, we explore whether the Nobel prize affects the credit share of Nobel laureates relative to their coauthors. Fig.~\ref{fig:evolution}a shows the evolution of credit share for the 1997 Nobel prize-winning paper in Physics~\cite{Chu1985}. We find that right after the publication Ashkin gets virtually all the credit for the discovery and Chu's credit share is tiny, given his lack of previous track record in this area (Fig.~\ref{fig:evolution}a). Yet, with time his credit share increases, while Ashkin's credit share decreases, partly because Ashkin stopped publishing papers after 1986 and retired in 1992. The method also helps us explore how the papers preceding the publication (i.e., previous reputation) of a prize-winning paper influence the credit allocation. Indeed, when we consider all co-cited papers, Ashkin's credit share is higher than the credit share of the laureate Chu, given his work preceding the 1985 paper. Yet, Chu gets higher credit share than Ashkin if we only consider the co-cited papers published after 1985 (Fig.~\ref{fig:evolution}a: inset). This example indicates that although established scientists receive more credit than their junior colleagues from their co-authored publications, this situation can change if the junior colleague makes important independent contribution to the field.

The effect of Nobel prize on credit share is also remarkable for the other 1997 Nobel prize-winning paper~\cite{Lett1988}: the Nobel laureate W. D. Phillips's credit share jumps after the prize year (Fig.~\ref{fig:evolution}b). Indeed, the prize, by canonizing credit, alters the subsequent citation patterns~\cite{Mazloumian2011,Ren2012,Ahn2010}, reflecting a ``rich get richer'' phenomenon in science~\cite{Merton1968,Barabasi1999,Albert2002,Garlaschelli2003,Petersen2011,Azoulay2014,Shen2014}. To quantify how widespread this effect is, we systematically studied how the credit share of laureates relative to their coauthors changes when they are awarded the Nobel prize. Therefore for each laureate we calculate her average credit share $c_b$ over $3$ years before the award year and the average credit share $c_a$ over 3 years after the award year. We quantify the increase of the credit share using the ratio $c_a / c_b$. For most Nobel laureates, the Nobel prize does improve their credit share relative to their coauthors (Fig.~\ref{fig:evolution}d), an effect that is the strongest in Physics and the weakest in Medicine. The two cases with the strongest effect in Physics (the outlier points in Fig.~\ref{fig:evolution}d) are shown in Fig.~\ref{fig:evolution}a,b.

\subsection{Comparing independent authors}
The developed method allows us to compare authors that are in the same research field but may not have published papers together. In this case, the co-citation strength is based on the citing papers which simultaneously cite at least one paper of each compared author, automatically identifying their common research topic (SI Appendix: Figure S1). Therefore the credit share of the compared authors reflects their relative contribution to their common research topic, just as the credit share quantifies the coauthors' relative contribution to a joint paper. An excellent example is provided by the 2013 Nobel prize in Physics (see SI Appendix: Figure S2 for more examples). The prize posed a widely publicized dilemma: six physicists and three key papers are credited for the 1964 discovery pertaining to the theory of Higgs boson, but the prize could be shared by a maximum of three individuals. F. Englert and R. Brout published the theory first~\cite{Englert1964} but failed to spell out the Higgs boson, whose existence was predicted in a subsequent paper by P. W. Higgs~\cite{Higgs1964}. G. S. Guralnik, C. R. Hagen, and T. W. B. Kibble, one month later proposed the same theory~\cite{Guralnik1964}, explaining how the building blocks of the universe get their mass. In 2010 the six physicists were given equal recognition by the American Physical Society (APS), sharing the Sakurai prize for theoretical particle physics. This symmetry was broken by the Nobel committee, awarding the prize to Higgs and Englert in 2013. To explore their credit share we apply our method to compare these researchers~(Fig.~\ref{fig:evolution}c), finding that Higgs gets the most credit, followed by Kibble, while Englert is the third, getting only slightly higher credit than his coauthor Brout (deceased). Finally, Guralnik and Hagen equally share the remaining credit. Therefore the scientific community assigns credit for the discovery recognized by the 2013 Nobel Physics prize to Higgs, Kibble, and Englert (and the deceased Brout), in this order. The committee, by bypassing Kibble, has clearly deviated from the community's perception of where the credit lies~\cite{BBCNews}.

\subsection{Robustness of the method}
We validate the robustness of our method by applying it to two disparate datasets, the publicly available APS dataset and the Web of Science (WOS) dataset (see SI Appendix: Section S1). The APS dataset consists of papers published by journals of APS between 1893 and 2009, hence its coverage is biased towards the US-based physics community~\cite{Redner2007}. The dataset does not contain papers published in interdisciplinary journals, like Science, where the 2010 Nobel winning paper was published. The WOS dataset, in contrast, contains all papers indexed by Thomson Reuters between 1955 and 2012~\cite{Wang2013}. By comparing the results obtained using these two datasets, we can evaluate the robustness of our method. In Table 1 we show the credit share of the papers obtained using each dataset individually. Overall, we find that the data incompleteness of the APS dataset never alters the relative ranking of the researchers. The same robustness is documented in Fig.~\ref{fig:evolution}c, where we calculated using both the APS dataset and the WOS dataset the dynamic credit share of the contenders for the 2013 Nobel prize, particularly important given that the European experimental particle physics community shuns the APS journals. Consequently, there are huge difference in coverage between the APS and WOS datasets. For example, for Higgs' prize winning paper we have $N_\mathcal{D}=187, N_\mathcal{P}=1,847$ in the APS dataset and $N_\mathcal{D}=879, N_\mathcal{P}=24, 596$ in the WOS dataset. Despite the bias of the APS dataset and the huge difference in coverage, the relative credit of the six authors remains unchanged (Fig.~\ref{fig:evolution}c).

\subsection{Exogenous information}
The proposed algorithm can incorporate exogenous information to improve its accuracy. To show this we explored five priors for constructing the credit allocation matrix ${\bf A}$, each reflecting a different hypothesis about the role of the authors. They are: (1) Count prior~\cite{Hirsch2005}: each author is viewed as the sole author of the particular publication; (2) Fractional prior~\cite{Hirsch2007}: authors equally share one credit independent of their position in the author list; (3) Harmonic prior~\cite{Hagen2008}: authors share one credit with their credit share proportional to the reciprocal of authors' rank in the author list; (4) Axiomatic prior~\cite{Stallings2013}: authors share one credit but the credit share of each author is determined by the number of coauthors with lower rank in the author list; and (5) Zhang's prior~\cite{Zhang2009}: the first and the corresponding authors get one credit while other authors share one credit dependent on their rank in the author list (see SI Appendix: Section S2.2 for details). The first two priors do not depend on the order of authors while the last three do. We summarize the results of each prior separately for three Nobel-awarding disciplines (SI Appendix: Table S3). We find that when we incorporate corresponding author information (if not available, we take the last author as the corresponding author), for Medicine and Chemistry the accuracy increases but drops for Physics. Therefore, if contextual or exogenous information is available, our method can absorb that, improving its predictive power. Yet, these other priors should be only used in a disciplinary fashion.

\section{Discussion}

In this paper we proposed a method to quantify the credit share of coauthors by reproducing the collective credit allocation process informally used by the scientific community. The method captures several key aspects of credit allocation in science: (1) Credit is allocated among scientists based on their perceived contribution rather than their actual contribution. (2) Established scientists receive more credit than their junior collaborators from coauthored publications~\cite{Merton1968}. This balance can change, however, if the junior colleague makes important independent contribution to the field. (3) Credit share changes with the evolution of the field.

Our method has several distinguishing characteristics, differentiating it from current credit allocation procedures that are based on the author list~\cite{Egghe2006,Hirsch2007,Hagen2008,Zhang2009,Stallings2013}: (1) The method offers topic-dependent credit share, as each paper's research topic is automatically defined by the body of papers that cite it. (2) It performs consistently better than existing methods across disciplines. Indeed, previous methods that assign credit to the first or the corresponding authors work only for disciplines that have clear agreed-upon rule on authorship and credit allocation. (3) The method is flexible, being able to incorporate the order of coauthors in the author list, allowing us to construct a credit allocation matrix that captures exogenous information (see SI Appendix: Section S2.2). (4) The method provides a natural way to directly compare the relative scientific impact of researchers that did not collaborate with each other but work in the same research field. (5) The method could be employed to refine some established measures for scientific impact by considering the credit share of coauthors. (6) As further improvement, we could consider a page-rank style algorithm, where the citing set $\mathcal{P}$ is weighted based on their citation count. Hence citations from more influential papers would gain more weight.

The proposed credit allocation method is based on citations, the most elementary form of visibility and credit in the scientific community. Consequently it does not explicitly account for other tokens of impact, like invited talks, keynotes, mentoring, books, each of which can alter the reputation of a scientist relative to its coauthors. However, our method may implicitly incorporate these effects: if these activities enhance an author's visibility compared to his/her coauthors, it could result in long-term changes in citations and credit share that are captured by the proposed method.

Finally, credit allocation has potential long-term impact on the career of individuals, affecting hiring, funding or promotion decisions. We wish to clarify that our algorithm does not capture the precise role of an individual in a paper or a discovery --- it only captures the community's \textit{perception} of each individual's contribution, as reflected by their body of work. Hence we would caution turning this algorithm into the sole tool for credit allocation --- letters from coauthors could offer a more nuanced or altogether different picture. Hence it should be used in conjunction with the other available evaluation tools. The method may also offer feedback to an individual of the need to seek ways to strengthen the credit for a work. It may also have adverse effects: uncovering the mechanism of credit allocation may increase the likelihood that some authors can ``jockey'' for position, seeking to change the outcome. However, such credit manipulation may be realistic only for lower impact work, where collective effects do not dominate the citation count. Finally, we must keep in mind that the algorithm relies on citation patterns that take time to accumulate. Hence young scientists, with fewer citations, no matter how important their contribution is, will be at disadvantage. We therefore must learn to account for age and time-dependent factors in credit allocation, opening up avenues for further research.





\begin{acknowledgments}
We thank Dashun Wang, Chaoming Song, Roberta Sinatra, Santiago Gil, Pierre Deville, Qing Jin, Burcu Yucesoy, Elena Renda, and all other colleagues at Center for Complex Network Research for valuable discussions and comments. HWS is supported by the National Basic Research Program of China (973 Program) under grant number 2014CB340401, the National High-tech R\&D Program of China (863 Program) under grant number 2014AA015103, and the National Natural Science Foundation of China (Nos. 61202215, 61232010). ALB is supported by Lockheed Martin Corporation (SRA 11.18.11), the Network Science Collaborative Technology Alliance is sponsored by the U.S. Army Research Laboratory under agreement W911NF-09-2-0053, Defence Advanced Research Projects Agency under agreement 11645021, the Future and Emerging Technologies Project 317 532 ``Multiplex'' financed by the European Commission, and the NIH, Centers of Excellence of Genomic Science (CEGS) under agreement 1P50HG4233.
\end{acknowledgments}


\begin{thebibliography}{40}


\bibitem{Wuchty2007}
Wuchty S, Jones BF, Uzzi B (2007)
The increasing dominance of teams in production of knowledge.
{\it Science} 316(5827):1036--1039.

\bibitem{Lawrence2007}
Lawrence PA (2007)
The mismeasurement of science.
{\it Curr Biol} 17(15):R583--R585.

\bibitem{Hodge1981}
Hodge SE, Greenberg DA (1981)
Publication credit.
{\it Science} 213(4511):950--950.

\bibitem{Kennedy2003}
Kennedy D (2003)
Multiple authors, multiple problems.
{\it Science} 301(5634):733--733.

\bibitem{Allen2014}
Allen L, Scott J, Brand A, Hlava M, Altman M (2014)
Credit where credit is due.
{\it Nature} 508(7496): 312--313.

\bibitem{Sekercioglu2008}
Sekercioglu CH (2008)
Quantifying coauthor contributions.
{\it Science} 322(5900):371--371.

\bibitem{Greene2007}
Greene M (2007)
The demise of the lone author.
{\it Nature} 450(7173):1165--1165.

\bibitem{Biggs2008}
Biggs J (2008)
Allocating the credit in collaborative research.
{\it Political Science \& Politics} 41(1):246--247.

\bibitem{Kaur2013}
Kaur J, Radicchi F, Menczer F (2013)
Universality of scholarly impact metrics.
{\it Journal of Informetrics} 7(4):924--932.

\bibitem{Lehmann2006}
Lehmann S, Jackson AD, Lautrup BE (2006)
Measures for measures.
Nature 444(7122):1003--1004.

\bibitem{Garfield1972}
Garfield E (1972)
Citation analysis as a tool in journal evaluation.
{\it Science} 178(4060):471--479.

\bibitem{Hirsch2005}
Hirsch JE (2005)
An index to quantify an individual's scientific research output.
{\it Proc Natl Acad Sci USA} 102(46):16569--16572.

\bibitem{Egghe2006}
Egghe L (2006)
Theory and practise of the g-index.
{\it Scientometrics} 69(1):131--152.

\bibitem{Hirsch2007}
Hirsch JE (2007)
Does the {H} index have predictive power?
{\it Proc Natl Acad Sci USA} 104(49):19193--19198.

\bibitem{Hagen2008}
Hagen NT (2008)
Harmonic allocation of authorship credit: source-level correction of bibliometric bias assures accurate publication and citation analysis.
{\it PLoS ONE} 3(12):e4021.

\bibitem{Zhang2009}
Zhang CT (2009)
A proposal for calculating weighted citations based on author rank.
{\it EMBO Rep} 10(5):416--417.

\bibitem{Stallings2013}
Stallings J, Vance E, Yang J, Vannier MW, Liang J, Pang L, Dai L,
  Ye I, Wang G (2013)
Determining scientific impact using a collaboration index.
{\it Proc Natl Acad Sci USA} 110(24):9680--9685.

\bibitem{Tscharntke2007}
Tscharntke T, Hochberg ME, Rand TA, Resh VH, Krauss J (2007)
Author sequence and credit for contributions in multiauthored
  publications.
{\it PLoS Biol} 5(1):e18.

\bibitem{Clippel2008}
De Clippel G, Moulin H, Tideman N (2008)
Impartial division of a dollar.
{\it J Econ Theor} 139(1):176--191.

\bibitem{Foulkes1996}
Foulkes W, Neylon N (1996)
Redefining authorship - {R}elative contribution should be given after each author's name.
{\it BMJ} 312(7043):1423--1423.

\bibitem{Campbell1999}
Campbell P (1999)
Policy on papers' contributors.
{\it Nature} 399(6735):393--393.

\bibitem{Radicchi2009}
Radicchi F, Fortunato S, Markines B, Vespignani A (2009)
Diffusion of scientific credits and the ranking of scientists.
{\it Phys Rev E Stat Nonlin Soft Matter Phys} 80(5 Pt 2):056103.

\bibitem{Rybski2009}
Rybski D, Buldyrev SV, Havlin S, Liljeros F, Makse HA (2009)
Scaling laws of human interaction activity.
{\it Proc Natl Acad Sci USA} 106(31):12640--12645.

\bibitem{Mones2014}
Mones E, Pollner P, Vicsek T (2014)
Universal hierarchical behavior of citation networks.
{\it arXiv}: 1401.4676.

\bibitem{Brune1996}
Brune M, Hagley E, Dreyer J, Ma\^{\i}tre X, Maali A, Wunderlich C, Raimond JM, Haroche S (1996)
Observing the progressive decoherence of the ¡°meter¡± in a quantum
  measurement.
{\it Phys Rev Lett} 77(24):4887--4890.

\bibitem{Meekhof1996}
Meekhof DM, Monroe C, King BE, Itano WM, Wineland DJ (1996)
Generation of nonclassical motional states of a trapped atom.
{\it Phys Rev Lett} 76(11):1796--1799.

\bibitem{Baibich1988}
Baibich MN, Broto JM, Fert A, Dau FNV, Petroff F, Etienne P,
  Creuzet G, Friederich A, Chazelas J (1988)
Giant magnetoresistance of {(001)Fe/(001)Cr} magnetic superlattices.
{\it Phys Rev Lett} 61(21):2472--2475.

\bibitem{Grunberg1986}
Gr{\"{u}}nberg P, Schreiber R, Pang Y, Brodsky MB, Sowers H (1986)
Layered magnetic structures: evidence for antiferromagnetic coupling
  of {Fe} layers across {Cr} interlayers.
{\it Phys Rev Lett} 57(19):2442--2445.

\bibitem{Small1973}
Small H (1973)
Co-citation in the scientific literature: a new measure of the
  relationship between two documents.
{\it J Am Soc Inf Sci} 24(4):265--269.

\bibitem{Novoselov2004}
Novoselov KS, Geim AK, Morozov SV, Jiang D, Zhang Y, Dubonos SV,
  Grigorieva IV, Firsov AA (2004)
Electric field effect in atomically thin carbon films.
{\it Science} 306(5696):666--669.

\bibitem{Hewish1968}
Hewish A, Bell SJ, Pilkingt JD, Scott PF, Collins RA (1968)
Observation of a rapidly pulsating radio source.
{\it Nature} 217(5130):709--713.

\bibitem{Chu1985}
Chu S, Hollberg L, Bjorkholm JE, Cable A, Ashkin A (1985)
Three-dimensional viscous confinement and cooling of atoms by
  resonance radiation pressure.
{\it Phys Rev Lett} 55(1):48--51.

\bibitem{McGloin2010}
McGloin D, Reid JP (2010)
Forty years of optical manipulation.
{\it Optics and Photonics News} 21(3):20--26.

\bibitem{Ashkin1970}
Ashkin A (1970)
Acceleration and trapping of particles by radiation pressure.
{\it Phys Rev Lett} 24(4):156--159.

\bibitem{Gordon1980}
Gordon JP, Ashkin A (1980)
Motion of atoms in a radiation trap.
{\it Phys Rev A} 21(5):1606--1617.


\bibitem{Lett1988}
Lett PD, Watts RN, Westbrook CI, Phillips WD, Gould PL, Metcalf HJ (1988)
Observation of atoms laser cooled below the doppler limit.
{\it Phys Rev Lett} 61(2):169--172.

\bibitem{Mazloumian2011}
Mazloumian A, Eom YH, Helbing D, Lozano S, Fortunato S (2011)
How citation boosts promote scientific paradigm shifts and {N}obel prizes.
{\it PLoS ONE} 6(5):e18975.

\bibitem{Ren2012}
Ren FX, Shen HW, Cheng XQ (2012)
Modeling the clustering in citation networks.
{\it Physica A} 391(12):3533--3539.

\bibitem{Ahn2010}
Ahn YY, Bagrow JP, Lehmann S (2010)
Link communities reveal multiscale complexity in networks.
{\it Nature} 466(7307):761--764.

\bibitem{Merton1968}
Merton RK (1968)
The {M}atthew effect in science.
{\it Science} 159(3810):56--63.

\bibitem{Barabasi1999}
Barab\'{a}si AL, Albert R (1999)
Emergence of scaling in random networks.
{\it Science} 286(5439):509--512, 1999.

\bibitem{Albert2002}
Albert R, Barab\'{a}si AL (2002)
Statistical mechanics of complex networks.
{\it Rev Mod Phys} 74(1): 47-97.

\bibitem{Garlaschelli2003}
Garlaschelli D, Caldarelli G, Pietronero L (2003)
Universal scaling relations in food webs.
{\it Nature} 423(6936):165--168.


\bibitem{Petersen2011}
Petersen AM, Jung WS, Yang JS, Stanley HE (2011)
Quantitative and empirical demonstration of the Matthew effect in a study of career longevity.
{\it Proc Natl Acad Sci USA} 108(1):18--23.


\bibitem{Azoulay2014}
Azoulay P, Stuart T, Wang Y (2014)
Matthew: Effect of Fable?
{\it Management Science} 60(1):92--109.

\bibitem{Shen2014}
Shen HW, Wang D, Song C, Barab\'asi AL (2014)
Modeling and predicting popularity dynamics via reinforced Poisson processes.
{\it Proceedings of the 28th AAAI Conference on Artificial Intelligence (AAAI Press, Palo Alto, California)}, pp 291--297.

\bibitem{Englert1964}
Englert F, Brout R (1964)
Broken symmetry and the mass of gauge vector mesons.
{\it Phys Rev Lett} 13(9):321--323.

\bibitem{Higgs1964}
Higgs PW (1964)
Broken symmetries and the masses of gauge bosons.
{\it Phys Rev Lett} 13(16):508--509.

\bibitem{Guralnik1964}
Guralnik GS, Hagen CR, Kibble TWB (1964)
Global conservation laws and massless particles.
{\it Phys Rev Lett} 13(20):585--587.

\bibitem{BBCNews}
BBC News (2014)
Early night cost Higgs credit for big physics theory. Avaiable at
http://m.bbc.co.uk/news/science-environment-26014584. Accessed February 18, 2014.

\bibitem{Redner2007}
Redner S (2007)
Citation statistics from 110 years of {P}hysical {R}eview.
{\it Phys. Today} 58(6):49-54.

\bibitem{Wang2013}
Wang D, Song C, Barab\'asi AL (2013)
Quantifying long-term scientific impact.
{\it Science} 342(6154):127--132.



\end{thebibliography}




\end{article}






\begin{table}
\caption{Credit share for five Nobel prize winning papers. \textbf{Credit share is computed according to Eq.~[\ref{eq:credit_share}] or Eq.~[\ref{eq:matrix_form}] in the awarding year of each paper, using the WOS and the APS datasets. Coauthors are shown according to their positions in the author list. The maximum credit share is highlighted in bold and the laureates are marked with asterisks. For papers not contained in the dataset, we put ``N/A'' for credit share. }}
\label{tab:example}
\begin{center}
\begin{tabular}{p{3cm}lcc}
\hline
\multirow{2}{3cm}{\textbf{Awarding Year / Paper}} & \multirow{2}{*}{\textbf{Authors}}
& \multicolumn{2}{c}{\textbf{Credit Share}}\\
\cline{3-4}
& & WOS & APS \\
\hline
\multirow{8}{3cm}{2012 / Phys. Rev. Lett. 77, 4887 (1996)} & M. Brune & 0.204 & 0.209 \\
& E. Hagley & 0.074 & 0.080 \\
& J. Dreyer & 0.065 & 0.070 \\
& X. Ma\^{\i}tre & 0.068 & 0.074 \\
& A. Maali & 0.073 & 0.077 \\
& C. Wunderlich & 0.069 & 0.074 \\
& J. M. Raimond & 0.212 & 0.206 \\
& S. Haroche$^*$ & \textbf{0.236} & \textbf{0.211} \\
\hline
\multirow{5}{3cm}{2012 / Phys. Rev. Lett. 76, 1796 (1996)}  &  D. M. Meekhof & 0.160 & 0.149 \\
& C. Monroe & 0.198 & 0.182 \\
& B. E. King & 0.173 & 0.158 \\
& W. M. Itano & 0.200 & 0.239 \\
& D. J. Wineland$^*$ & \textbf{0.270} & \textbf{0.272}  \\
\hline
\multirow{8}{3cm}{2010 / Science 306, 666 (2004)}  &  K. S. Novoselov$^*$ & \textbf{0.244} & N/A \\
& A. K. Geim$^*$ & \textbf{0.253} & N/A  \\
& S. V. Morozov & 0.111 & N/A \\
& D. Jiang & 0.102 & N/A \\
& Y. Zhang & 0.064 & N/A \\
& S. V. Dubonos & 0.075 & N/A \\
& I. V. Grigorieva & 0.075 & N/A \\
& A. A. Firsov & 0.075 & N/A \\
\hline
\multirow{9}{3cm}{2007 / Phys. Rev. Lett. 61, 2472 (1988)}  & M. N. Baibich & 0.094 & 0.093 \\
& J. M. Broto & 0.090 & 0.090 \\
& A. Fert$^*$ & \textbf{0.242} & \textbf{0.252} \\
& F. Nguyen Van Dau & 0.093 & 0.093 \\
& F. Petroff & 0.114 & 0.100 \\
& P. Etienne & 0.089 & 0.093 \\
& G. Creuzet & 0.097 & 0.093 \\
& A. Friederich & 0.091 & 0.093 \\
& J. Chazelas & 0.090 & 0.093 \\
\hline
\multirow{5}{3cm}{1997 / Phys. Rev. Lett. 55, 48 (1985)}  & S. Chu$^*$ & 0.244 & 0.196 \\
& L. Hollberg & 0.087 & 0.096 \\
& J. E. Bjorkholm & 0.134 & 0.162 \\
& A. Cable & 0.138 & 0.160 \\
& A. Ashkin & \textbf{0.397} & \textbf{0.386} \\
\hline
\end{tabular}
\end{center}
\end{table}



\end{document}